\documentclass[eqsecnum,superscriptaddress,nofootinbib,11pt]{revtex4}
\usepackage{amsmath}
   \usepackage{bm}
\usepackage{amsthm,amscd}
\usepackage{mathrsfs}
\usepackage{verbatim}
\usepackage{amsfonts,amsmath,latexsym,amssymb,txfonts,bm}
\usepackage[american]{babel}
\usepackage{dsfont}
\usepackage[dvips]{graphicx}
\usepackage{latexsym}
\usepackage{epsfig}
\usepackage{braket}
\usepackage{color}
\newcommand{\diff}{\mathrm{d}}

\def\beq{\begin{eqnarray}}
\def\eeq{\end{eqnarray}}

\begin{document}
\title{Vacuum instability in Kaluza-Klein manifolds}

\author{Guglielmo Fucci\footnote{Electronic address: fuccig@ecu.edu}}
\affiliation{Department of Mathematics, East Carolina University, Greenville, NC 27858 USA}


\date{\today}
\vspace{2cm}
\begin{abstract}

The purpose of this work in to analyze particle creation in spaces with extra dimensions. We consider, in particular, a massive scalar
field propagating in a Kaluza-Klein manifold subject to a constant electric field. We compute the rate of particle creation from vacuum 
by using techniques rooted in the spectral zeta function formalism. The results we obtain show explicitly how the presence
of the extra-dimensions and their specific geometric characteristics, influence the rate at which pairs of particles and anti-particles are generated.

\end{abstract}
\maketitle

\section{Introduction}

In 1936 Heisenberg, together with his student Euler, predicted for the first time the astonishing phenomenon of vacuum instability  \cite{heisen36}.
Some time later, in 1951, by exploiting field theoretical methods, Julian Schwinger described the phenomenon in great detail within the ambit of
quantum electrodynamics \cite{schwing51}. In his seminal work, he computed, in particular, the rate of particle creation for massive spinless
fields under the influence of a constant electric field. As a tribute for his pioneering work, particle creation due to the presence of an external
electric field is also known as {\it Schwinger's mechanism}. Since then, vacuum instability has been the main focus of a large number of works who
analyze the phenomenon in a variety of different cases (for a review see e.g. \cite{dunne05,gelis15}).
Although vacuum instability due to an external field is now a fairly well-understood subject \cite{gavrilov95}, its direct experimental verification remains
still elusive due to the prohibitively large electric fields needed for particle creation \cite{dunne09}.

In order to overcome this difficulty, a number of methods have been proposed and analyzed in the literature which could enhance the
rate of particle creation. Among these are the use of high energy lasers and shaped laser pulses to achieve strong enough electric fields \cite{dumlu10,dunne09,heben09} and
dynamically assisted mechanisms which rely on the theoretical observation that by superimposing different time-dependent electric fields particle creation can be enhanced
\cite{brezin70,orth11,schu08}. All these methods have in common the concept that by generalizing the type of electric fields one considers, then the energy threshold necessary
for particle creation can more easily be achieved (for the purpose of experimental observation of Schwinger's mechanism).
Although exploiting general and time-dependent electric fields might be the most experimentally viable method to enhance and possibly observe particle creation
from vacuum, there exist other approaches aimed at the same objective. In fact, the pair production rate due to a constant electric field can be increased, for instance, by the presence of a gravitational field \cite{avramidi09,fucci10}. In this work we would like to investigate whether the presence of extra dimensions
can intensify the rate of particle creations from vacuum. This question is part of the larger issue of understanding how topological changes to the
underlying space-time influence the vacuum instability and whether particular geometric configurations could be exploited with the specific purpose of
enhancing the particle creation to the point of being more easily detectable. Space-time manifolds that have a more general structure than the flat Minkowski
space have, in the past few decades, become extremely relevant in fundamental physics. In fact, in string theory it is assumed that
the Universe consists of a higher-dimensional manifold of which only four dimensions are observable \cite{bb}. The remaining dimensions form a compact manifold whose
presence should influence physical phenomena in the observable four-dimensional space.
The comment offered in the last sentence provides the main motivation for the present analysis which consists in a detailed investigation into how, and to what extent, the
presence of extra, compactified, dimensions influence the rate of particle creations due to an external homogeneous electric field.
We would like to point out that the influence the extra dimensions have on our four-dimensional Universe has been studied by many authors in a variety of
different settings. For instance, the effects of extra dimensions have been analyzed in the ambit of Bose-Einstein condensation \cite{fucci11},
the Casimir effect (see e.g. \cite{elizalde09,kirsten09,pop}), and cosmology \cite{deff,giudice,rub}. It is important to emphasize that these works
provide a small, and by no means inclusive, set of examples which focus on the influence of extra dimensions and that several
other theoretically and physically relevant settings have been considered in the literature.

In this work we utilize spectral zeta function techniques to compute the one-loop effective action for a massive scalar field propagating in a
higher-dimensional space and subject to a constant electric field.  From the expression of the one-loop effective action we then extract its
imaginary part which describes the rate of particle creation. The main goal is to present explicit results for the rate of particle creation
and their dependence on the geometric characteristics of the extra dimensions. In this paper we work in units such that $\hbar=c=e=1$.

The outline of the paper is as follows. In the next section we utilize spectral zeta function regularization techniques in order to
write the one-loop effective action for a massive scalar field under the influence of an homogeneous electric field. Section \ref{Sec3}  focuses on the explicit evaluation
of the imaginary part of the derivative at $s=0$ of the spectral zeta function. The calculations rely heavily on known results regarding the
imaginary part of the Hurwitz zeta function of imaginary second argument. In Section \ref{Sec4} we find the rate of particle creation on product manifolds
and we specialize the general result to a few simple particular cases. The conclusions summarize the main results of this work and outline a few areas
in which further research on the topic can be performed.

\section{One-loop effective action and the spectral zeta function}

We consider a $D$-dimensional, $D=d+4$, product manifold of the type ${\cal M}=M_{4}\times N$, where $M_{4}$ denotes the usual four-dimensional Minkowski
space-time and $N$ is a $d$-dimensional compact, smooth Riemannian manifold with or without boundary. As previously mentioned, we focus our analysis on
massive scalar fields under the influence of an external electric field. The relevant differential operator describing
the dynamics of this system is
\begin{equation}\label{0}
  {\cal L}=-g^{\mu\nu}{\cal D}_{\mu}{\cal D}_{\nu}+m^2\;,
\end{equation}
where $m>0$ denotes the mass of the field and ${\cal D}_{\mu}=\nabla_{\mu}+ i A_{\mu}$ is the covariant derivative with respect to the $U(1)$ connection
with $A_{\mu}$ denoting the vector potential. We assume that the electromagnetic field $F_{\mu\nu}$ possesses only one non-vanishing invariant $2I=F^{\mu\nu}F_{\mu\nu}<0$.
Under this assumption, there exists a reference frame in which only $F_{0i}\neq 0$. This situation corresponds to an electromagnetic field which is purely electric \cite{gavrilov95}.
We further suppose that the electric field propagates in only, say, the $z$ direction. In our set up, the $D$-dimensional vector potential can be written as $A_{\mu}=(0,0,0,Ex^0,0,\ldots,0)$
where $E$ denotes the electric field strength. The operator ${\cal L}$ acting on suitable scalar functions defined on ${\cal M}$ can, hence, be explicitly written as
\begin{equation}\label{1}
  {\cal L}=\partial_{x^{0}}^{2}-\partial_{x^{1}}^{2}-\partial_{x^{2}}^{2}-\left(\partial_{x^{3}}-i Ex^0\right)^{2}+\Delta_{N}+m^2\;,
\end{equation}
where $\Delta_{N}$ represents the Laplace operator on the manifold $N$. In order to obtain a spectral problem for which the zeta function is well-defined,
it is necessary to perform a Wick rotation $x^0\to i\tau$ to imaginary time \cite{dewitt}. After performing the rotation, the Minkowski space $M_{4}$ becomes Euclidean, $E_{4}$, and
${\cal L}$ becomes an elliptic operator $\mathscr{L}$ acting on functions on $\overline{{\cal M}}=E_{4}\times N$.

We are interested in the following eigenvalue problem
\begin{equation}\label{2}
  \mathscr{ L} \phi_{i}({\bf x},X)=\lambda_{i}\phi_{i}({\bf x},X)\;,
\end{equation}
the solution of which can be found by separation of variables. By noticing that $x^{1}$ and $x^{2}$ are two unbounded directions, and by denoting by $\varphi_{j}$ the
eigenfunctions of $\Delta_{N}$ corresponding to the eigenvalues $\alpha_{i}$
\begin{equation}\label{3}
  -\Delta_{N}\varphi_{j}(X)=\alpha^2_{j}\varphi_{j}(X)\;,
\end{equation}
we can write the eigenfunctions $\phi_{i}$ as
\begin{equation}\label{4}
  \phi_{i}({\bf x},X)=e^{i{\bf k}\cdot {\bf x}}\varphi_{i}(X)f(x^0)\;,
\end{equation}
where the momentum ${\bf k}=(k_{1},k_{2},k_{3})\in\mathbb{R}^{3}$, and $f(x^0)$ satisfies the one-dimensional harmonic oscillator equation
\begin{equation}\label{5}
  \left[-\partial_{0}^{2}+\left(k_{3}-Ex^{0}\right)^2\right]f(x^0)=\omega f(x^0)\;,
\end{equation}
with the eigenvalues $\omega=E(2n+1)+k_{1}^{2}+k_{2}^{2}$, $n\in\mathbb{N}_{0}$. By collecting the above results it is not very difficult to find
that the desired eigenvalues $\lambda$ of the elliptic operator $\mathscr{L}$ are
\begin{equation}\label{6}
  \lambda_{n,i}=\alpha^{2}_{i}+E(2n+1)+k_{1}^{2}+k_{2}^{2}+m^{2}\;.
\end{equation}

Since the eigenvalues $\lambda_{n,i}$ do not depend on the momentum $k_{3}$, each of them is degenerate with continuous
multiplicity $d(k_{3})$. In this case the spectral zeta function associated with $\mathscr{L}$ can be written as
\begin{equation}\label{7}
  \overline{\zeta}(s)=\sum_{i}d(\alpha)\sum_{n=0}^{\infty}\frac{d(k_{3})L_{1}L_{2}}{4\pi^2}\int_{-\infty}^{\infty}\int_{-\infty}^{\infty}\left[\alpha^{2}_{i}+E(2n+1)+k_{1}^{2}+k_{2}^{2}+m^{2}\right]^{-s}\diff k_{1}\diff k_{2}\;,
\end{equation}
where $d(\alpha)$ denotes the degeneracy of the eigenvalues $\alpha^{2}_{i}$, and $L_{1}$ and $L_{2}$ are unit lengths along the direction $x_{1}$, respectively, $x_{2}$. In order to obtain an expression for the degeneracy $d(k_{3})$
we exploit a method based on the small-$t$ asymptotic expansion of the heat kernel (see e.g. \cite{blau91}).
By using the eigenvalues in (\ref{6}) one can construct the trace of the heat kernel associated with the operator $\mathscr{L}$ as
\begin{equation}\label{8}
  \textrm{k}(t)=\frac{L_{1}L_{2}}{4\pi^2}\sum_{i}d(\alpha)\sum_{n=0}^{\infty}d(k_{3})\int_{-\infty}^{\infty}\int_{-\infty}^{\infty}e^{-t\left(k_{1}^{2}+k_{2}^{2}\right)}
  e^{-t\left[\alpha^{2}_{i}+E(2n+1)+m^2\right]}\diff k_{1}\diff k_{2}\;.
\end{equation}
Once the elementary integration over the unconstrained momenta is performed, we obtain
\begin{equation}\label{9}
\textrm{k}(t)=\frac{L_{1}L_{2}}{4\pi^2 t}K_{N}(t)\sum_{n=0}^{\infty}d(k_{3})e^{-t\left[E(2n+1)+m^2\right]}\;,
\end{equation}
where $K_{N}(t)$ denotes the trace of the heat kernel associated with the Laplacian $\Delta_{N}$ on the manifold $N$. By exploiting the geometric series, the remaining sum can be easily computed leading to the final result
\begin{equation}\label{10}
  \textrm{k}(t)=\frac{L_{1}L_{2}d(k_{3})}{8\pi^2 t}K_{N}(t)\frac{e^{-tm^{2}}}{\sinh(tE)}\;.
\end{equation}
According to the general theory, the trace of the heat kernel associated with the elliptic operator $\mathscr{L}$ on the manifold $E_{4}\times N$ has a small-$t$ asymptotic expansion
with leading term of the form \cite{gilkey95,kirsten01,vassi03}
\begin{equation}\label{11}
  \textrm{k}(t)\sim\frac{{\cal V}\,\textrm{Vol}(N)}{(4\pi t)^{2+\frac{d}{2}}}\;,
\end{equation}
where ${\cal V}$ denotes the \textit{unit} volume of the Euclidean space $E_{4}$. By exploiting a similar small-$t$
asymptotic expansion of $K_{N}(t)$, we find that, as $t\to 0$, $\textrm{k}(t)$ in (\ref{10}) reduces to
\begin{equation}\label{12}
\textrm{k}(t)\sim \frac{2\pi L_{1}L_{2}d(k_{3})\textrm{Vol}(N)}{(4\pi t)^{2+\frac{d}{2}}E}\;.
\end{equation}
By comparing (\ref{11}) and (\ref{12}) it is not difficult to obtain the following expression for the degeneracy
\begin{equation}\label{13}
  d(k_{3})=2\left(\frac{SE}{2\pi}\right)=\frac{SE}{\pi}\;,
\end{equation}
where $S$ is a two-dimensional unit volume and the factor of $2$ has been introduced in the degeneracy in order to account for both particles and antiparticles.

By substituting (\ref{13}) in (\ref{7}), and by performing the integration over the variables $k_{1}$ and $k_{2}$ we finally obtain the desired expression for the spectral zeta function
\begin{equation}\label{14}
\overline{\zeta}(s)=\frac{{\cal V} E}{4\pi^2(s-1)}\sum_{i}d(\alpha)\sum_{n=0}^{\infty}\left[\alpha^{2}_{i}+E(2n+1)+m^{2}\right]^{-s+1}
\end{equation}
which is well defined for $\Re s>D/2$. In order to compute the derivative of the spectral zeta function
at $s=0$, we need to analytically continue (\ref{14}) to a meromorphic function of $s$ in the entire complex plane.
To this end, we use, for $\Re s>D/2$, the Mellin-Barnes integral representation \cite{paris} for the zeta function in (\ref{14}), to get
\begin{equation}\label{15}
  \overline{\zeta}(s)=\frac{{\cal V} E}{4\pi^2\Gamma(s)}\int_{0}^{\infty}t^{s-2}\sum_{n=0}^{\infty}e^{-[E(2n+1)+m^{2}]t}\sum_{i}d(\alpha)e^{-\alpha_{i}^{2}t}\diff t\;.
\end{equation}
By assuming that the quantity $E(2n+1)+m^{2}$, $n\in\mathbb{N}_{0}$, is large, which is certainly true for the kinds of electric fields one
considers in the ambit of pair creation, we can utilize the well-known small-$t$ asymptotic expansion of the
of the trace of the heat kernel associated with the Laplacian on $N$ ,
\begin{equation}\label{16}
K_{N}(t)=\sum_{i}d(\alpha)e^{-\alpha_{i}^{2}t}\sim \frac{1}{(4\pi t)^{\frac{d}{2}}}\sum_{k=0}^{\infty}A^{N}_{\frac{k}{2}}t^{\frac{k}{2}}\;,
\end{equation}
where $A_{k/2}^{N}$ are universal coefficients constructed from geometric invariants of the manifold $N$ \cite{gilkey95,kirsten01,vassi03},
and perform the integral in (\ref{15}) to obtain the expression, valid for $\Re s>D/2$,
\begin{equation}\label{17}
\overline{\zeta}(s)\sim \frac{4{\cal V} E}{(4\pi)^{2+\frac{d}{2}}}\frac{\Gamma\left(s+\frac{k-d}{2}-1\right)}{\Gamma(s)}\sum_{k=0}^{\infty}A^{N}_{\frac{k}{2}}\sum_{n=0}^{\infty}\left[E(2n+1)+m^{2}\right]^{-s-\frac{k-d}{2}+1}\;.
\end{equation}
The desired analytic continuation to $\Re s\leq D/2$ is obtained by noticing that the sum over the index $n$ can be expressed in terms of the Hurwitz zeta function \cite{erde81,whit90}
\begin{equation}\label{18}
\sum_{n=0}^{\infty}\left[E(2n+1)+m^{2}\right]^{-s-\frac{k-d}{2}+1}=(2E)^{-s-\frac{k-d}{2}+1}\zeta_{H}\left(s+\frac{k-d}{2}-1;\frac{1}{2}+\frac{m^{2}}{2E}\right)\;,
\end{equation}
to get
\begin{equation}\label{19}
\overline{\zeta}(s)\sim \frac{2(2E)^{-s-\frac{k-d}{2}+2}\Omega}{(4\pi)^{2+\frac{d}{2}}}\frac{\Gamma\left(s+\frac{k-d}{2}-1\right)}{\Gamma(s)}
\sum_{k=0}^{\infty}a_{\frac{k}{2}}^{N}(2E)^{-\frac{k}{2}}\zeta_{H}\left(s+\frac{k-d}{2}-1;\frac{1}{2}+\frac{m^{2}}{2E}\right)\;,
\end{equation}
where we have redefined the heat kernel coefficients as follows $a_{k/2}^{N}=A_{k/2}^{N}/\textrm{Vol}(N)$, so that $a_{0}^{N}=1$.
By performing, now, an inverse Wick rotation $E\to -iE$ back to Minkowski spacetime we find an expression for $\zeta(s)$ associated with the operator ${\cal L}$ in the physically
relevant space ${\cal M}$
\begin{equation}\label{20}
\zeta(s)\sim \frac{2\Omega(-2iE)^{-s-\frac{k-d}{2}+2}}{(4\pi)^{2+\frac{d}{2}}}\frac{\Gamma\left(s+\frac{k-d}{2}-1\right)}{\Gamma(s)}
\sum_{k=0}^{\infty}a_{\frac{k}{2}}^{N}(-2iE)^{-\frac{k}{2}}\zeta_{H}\left(s+\frac{k-d}{2}-1;\frac{1}{2}+i\frac{m^{2}}{2E}\right)\;.
\end{equation}

In quantum field theory, the in-out vacuum transition amplitude is described in terms of the one-loop effective action as
\begin{equation}
 \braket{\textrm{out}|\textrm{in}}=\exp[i\Gamma_{(1)}]\;.
\end{equation}
An isolated quantum system obeys the unitarity condition and hence the associated one-loop effective action is real. A quantum system
interacting with an external field, such as a uniform electric field, is no longer isolated and its one-loop effective action is allowed to have, in general, both a real and an imaginary
part. While the real part of the one-loop effective action describes the polarization of vacuum of the massive scalar field by the external electric field \cite{schwing51},
the imaginary part characterizes particle creation.
In fact, the probability of production of particles in the whole spacetime is
\begin{equation}
P=1-| \braket{\textrm{out}|\textrm{in}}|^{2}=1-e^{-2\Im\Gamma_{(1)}}\;.
\end{equation}
In the framework of spectral zeta function regularization, the one-loop effective action
is expressed as \cite{byt,elizalde94,haw77,kirsten01}
\begin{equation}
  \Gamma_{(1)}=\frac{\diff}{\diff s}\left[\frac{\mu^{-2s}}{2}\zeta(s)\right]\Bigg|_{s=0}\;,
\end{equation}
where $\mu$ is a parameter with the dimension of a mass. Therefore, for the rate of particle creation one has \cite{schwing51}
\begin{equation}\label{21}
  R=\frac{P}{\Omega} \approx \frac{2\Im \Gamma_{(1)}}{\Omega}=\frac{1}{\Omega}\Im\left\{\frac{\diff}{\diff s}\left[\mu^{-2s}\zeta(s)\right]\Big|_{s=0}\right\}\;,
\end{equation}
where we have used the fact that $\Im \Gamma_{(1)}$ is generally small.
It is clear from the last expression that our next task consists in explicitly computing the imaginary part of the derivative at $s=0$ of the
zeta function displayed in (\ref{20}).

\section{Imaginary part of the one-loop effective action}\label{Sec3}

In order to compute the imaginary part of $\zeta'(0)$, we rely mainly on the results obtained in \cite{fucci11a} for the imaginary part of
the Hurwitz zeta function of imaginary second argument. From the expression (\ref{20}) it is straightforward to get
\begin{eqnarray}\label{22}
 \frac{\diff}{\diff s}\left[\mu^{-2s}\zeta(s)\right]&\sim& \frac{2\Omega(-2iE)^{\frac{d}{2}+2}}{(4\pi)^{2+\frac{d}{2}}\Gamma(s)}\left(-i\frac{2E}{\mu^{2}}\right)^{-s}
\sum_{k=0}^{\infty}a_{\frac{k}{2}}^{N}(-2iE)^{-\frac{k}{2}}\Gamma\left(s+\frac{k-d}{2}-1\right)\nonumber\\
&\times&\Bigg\{\left[\Psi\left(s+\frac{k-d}{2}-1\right)-\Psi(s)-\ln\left(-i\frac{2E}{\mu^{2}}\right)\right]\zeta_{H}\left(s+\frac{k-d}{2}-1;\frac{1}{2}+i\frac{m^{2}}{2E}\right)\nonumber\\
&+&\zeta'_{H}\left(s+\frac{k-d}{2}-1;\frac{1}{2}+i\frac{m^{2}}{2E}\right)\Bigg\}\;.
\end{eqnarray}
The limit as $s\to 0$ can be more conveniently evaluated if we distinguish between even and odd dimension $d$. By exploiting the relations, valid for $n\in\mathbb{N}_{0}$
\begin{eqnarray}\label{23}
  \frac{\Gamma(s-n)}{\Gamma(s)}&=&\frac{(-1)^{n}}{n!}+O(s)\;,\quad \Psi(s-n)-\Psi(s)=H_{n}+O(s)\;,\quad \Psi\left(s\pm n\pm\frac{1}{2}\right)-\Psi(s)=\frac{1}{s}+O(s)\;,\;\;\;
  \end{eqnarray}
where $H_{n}$ denotes the $n$-th harmonic number, one can prove that for $d=2l$, $l\in\mathbb{N}^{+}$, the derivative in (\ref{22}), at $s=0$, takes the form
\begin{eqnarray}\label{24}
 \frac{1}{\Omega}\frac{\diff}{\diff s}\left[\mu^{-2s}\zeta(s)\right]\Big|_{s=0}\!\!\!&\sim& \frac{2(-2iE)^{l+2}}{(4\pi)^{l+2}}\sum_{k=0}^{l+1}a_{k}^{N}\frac{(-1)^{l+1-k}(-2iE)^{-k}}{(l+1-k)!}\Bigg\{
 \left[H_{l+1-k}-\ln\left(-i\frac{2E}{\mu^{2}}\right)\right]\zeta_{H}\left(k-l-1;\frac{1}{2}+i\frac{m^{2}}{2E}\right)\nonumber\\
 &+&\zeta'_{H}\left(k-l-1;\frac{1}{2}+i\frac{m^{2}}{2E}\right)\Bigg\}
 +\frac{2}{(4\pi)^{l+2}}a_{l+2}^{N}\left[\textrm{FP}\,\zeta_{H}\left(1;\frac{1}{2}+i\frac{m^{2}}{2E}\right)-\ln\left(-i\frac{2E}{\mu^{2}}\right)\right]\nonumber\\
 &+&\frac{2(-2iE)^{l+2}}{(4\pi)^{l+2}}\sum_{k=0}^{\infty}a_{k+\frac{1}{2}}^{N}(-2iE)^{-k-\frac{1}{2}}
 \Gamma\left(k-l-\frac{1}{2}\right)\zeta_{H}\left(k-l-\frac{1}{2};\frac{1}{2}+i\frac{m^{2}}{2E}\right)\nonumber\\
 &+&\frac{2(-2iE)^{l+2}}{(4\pi)^{l+2}}\sum_{k=l+3}^{\infty}a_{k}^{N}(-2iE)^{-k}
 \Gamma\left(k-l-1\right)\zeta_{H}\left(k-l-1;\frac{1}{2}+i\frac{m^{2}}{2E}\right)\;,
\end{eqnarray}
where $\textrm{FP}$ denotes the finite part.
When the dimension of the manifold $N$ is, instead, odd, namely $d=2l+1$, $l\in\mathbb{N}_{0}$, the results in (\ref{23}) are once again used in obtaining the formula
\begin{eqnarray}\label{25}
 \frac{1}{\Omega}\frac{\diff}{\diff s}\left[\mu^{-2s}\zeta(s)\right]\Big|_{s=0}\!\!\!&\sim& \frac{2(-2iE)^{l+\frac{5}{2}}}{(4\pi)^{l+\frac{5}{2}}}\sum_{k=0}^{l+1}a_{k+\frac{1}{2}}^{N}\frac{(-1)^{l+1-k}(-2iE)^{-k-\frac{1}{2}}}{(l+1-k)!}\Bigg\{
 \left[H_{l+1-k}-\ln\left(-i\frac{2E}{\mu^{2}}\right)\right]\nonumber\\
 &\times&\zeta_{H}\left(k-l-1;\frac{1}{2}+i\frac{m^{2}}{2E}\right)
 +\zeta'_{H}\left(k-l-1;\frac{1}{2}+i\frac{m^{2}}{2E}\right)\Bigg\}\nonumber\\
 &+&\frac{2}{(4\pi)^{l+\frac{5}{2}}}a_{l+\frac{5}{2}}^{N}\left[\textrm{FP}\,\zeta_{H}\left(1;\frac{1}{2}+i\frac{m^{2}}{2E}\right)-\ln\left(-i\frac{2E}{\mu^{2}}\right)\right]\nonumber\\
 &+&\frac{2(-2iE)^{l+\frac{5}{2}}}{(4\pi)^{l+\frac{5}{2}}}\sum_{k=0}^{\infty}a_{k}^{N}(-2iE)^{-k}
 \Gamma\left(k-l-\frac{3}{2}\right)\zeta_{H}\left(k-l-\frac{3}{2};\frac{1}{2}+i\frac{m^{2}}{2E}\right)\nonumber\\
 &+&\frac{2(-2iE)^{l+\frac{5}{2}}}{(4\pi)^{l+\frac{5}{2}}}\sum_{k=l+3}^{\infty}a_{k+\frac{1}{2}}^{N}(-2iE)^{-k-\frac{1}{2}}
 \Gamma\left(k-l-1\right)\zeta_{H}\left(k-l-1;\frac{1}{2}+i\frac{m^{2}}{2E}\right)\;.
\end{eqnarray}
We can, now, proceed with the explicit evaluation of the imaginary part of (\ref{24}) and (\ref{25}).

To better describe the necessary computations, it is convenient to consider each of the four terms composing (\ref{24}) separately.
These terms, in turn, are expressed as products of a purely real component and a complex-valued one. It is clear that the evaluation of the
imaginary part of the latter is sufficient for our purposes. Since for any real number $\alpha$ and any complex-valued function $F$ we have
\begin{equation}\label{25a}
  \Im[(-2iE)^{\alpha}F]=(2E)^{\alpha}\left[\cos\left(\frac{\pi\alpha}{2}\right)\Im F-\sin\left(\frac{\pi\alpha}{2}\right)\Re F\right]\;,
\end{equation}
we can write the imaginary part of the complex-valued element in the first term of (\ref{24}) as follows
\begin{eqnarray}\label{26}
 \lefteqn{\Im\Bigg[(-2iE)^{l+2-k}\Bigg\{
 \left[H_{l+1-k}-\ln\left(-i\frac{2E}{\mu^{2}}\right)\right]\zeta_{H}\left(k-l-1;\frac{1}{2}+i\frac{m^{2}}{2E}\right)+\zeta'_{H}\left(k-l-1;\frac{1}{2}+i\frac{m^{2}}{2E}\right)\Bigg\}\Bigg]}\nonumber\\
 &=&-(2E)^{2+l-k}\Bigg\{\sin\left(\frac{(l+1-k)\pi}{2}\right)\Im [{\cal A}_{l+1-k}(E)]+\cos\left(\frac{(l+1-k)\pi}{2}\right)\Re[{\cal A}_{l+1-k}(E)]\Bigg\}\;,
\end{eqnarray}
where we have defined, for typographical convenience, for $n\in\mathbb{N}_{0}$,
\begin{equation}\label{27}
  {\cal A}_{n}(E)=
 \left[H_{n}-\ln\left(-i\frac{2E}{\mu^{2}}\right)\right]\zeta_{H}\left(-n\,;\frac{1}{2}+i\frac{m^{2}}{2E}\right)+\zeta'_{H}\left(-n\,;\frac{1}{2}+i\frac{m^{2}}{2E}\right)\;.
\end{equation}
The presence of the trigonometric functions in (\ref{26}) suggests that it is suitable to distinguish between even and odd values of the quantity $(l+1-k)$.
Therefore, for $(l+1-k)=2p$, $p\in\mathbb{N}_{0}$ we can write
\begin{eqnarray}\label{29}
 && \Im\Bigg[(-2iE)^{2p+1}\Bigg\{
 \left[H_{2p}-\ln\left(-i\frac{2E}{\mu^{2}}\right)\right]\zeta_{H}\left(-2p\,;\frac{1}{2}+i\frac{m^{2}}{2E}\right)+\zeta'_{H}\left(-2p\,;\frac{1}{2}+i\frac{m^{2}}{2E}\right)\Bigg\}\Bigg]\nonumber\\
 &=&(-1)^{p+1}(2E)^{2p+1}\Re[{\cal A}_{2p}(E)]\;.
\end{eqnarray}
By exploiting the relation \cite{fucci11a}, valid for all $s\in\mathbb{C}$ and $q\in\mathbb{R}$,
\begin{equation}\label{29a}
\zeta_{H}\left(s;\frac{1}{2}+iq\right)=2^{s}\zeta_{H}(s;2iq)-\zeta_{H}(s;iq)\;,
\end{equation}
we can write $\Re[{\cal A}_{2p}(E)]$ in terms of Hurwitz zeta functions of purely imaginary second argument as follows
\begin{eqnarray}\label{30}
\Re[{\cal A}_{2p}(E)]&=&\left[2^{-2p}\Re\zeta_{H}\left(-2p\,;i\frac{m^{2}}{E}\right)-\Re\zeta_{H}\left(-2p\,;i\frac{m^{2}}{2E}\right)\right]\left[H_{2p}-\ln\left(\frac{2E}{\mu^{2}}\right)\right]\nonumber\\
&-&\frac{\pi}{2}\left[2^{-2p}\Im\zeta_{H}\left(-2p\,;i\frac{m^{2}}{E}\right)-\Im\zeta_{H}\left(-2p\,;i\frac{m^{2}}{2E}\right)\right]+2^{-2p}\ln 2\Re\zeta_{H}\left(-2p\,;i\frac{m^{2}}{E}\right)\nonumber\\
&+&2^{-2p}\Re\zeta'_{H}\left(-2p\,;i\frac{m^{2}}{E}\right)-\Re\zeta'_{H}\left(-2p\,;i\frac{m^{2}}{2E}\right)\;.
\end{eqnarray}
In the ambit of pair creation the electric fields considered are large enough so that
$E>m^2$ and the expression in (\ref{30}) can be simplified further thanks to the following formulas \cite{fucci11a}, valid for $p\in\mathbb{N}_{0}$ and $q\in(0,1)$,
\begin{equation}\label{31}
  \Re\zeta_{H}(-2p\,; iq)=\frac{(-1)^{p}}{2}q^{2p}\;,\quad \Im\zeta_{H}(-2p\,; iq)=(2p)!\sum_{j=0}^{p-1}\frac{(-1)^{j+1}B_{2(p-j)}}{[2(p-j)]!(2j+1)!}q^{2j+1}-\frac{(-1)^{p}}{2p+1}q^{2p+1}\;,
\end{equation}
and
\begin{equation}\label{32}
  \Re\zeta'_{H}(-2p\,; iq)=\frac{\pi}{2}\Im\zeta_{H}(-2p\,; iq)-\frac{(-1)^{p}}{2}q^{2p}\ln q+(-1)^{p}\frac{(2p)!}{2(2\pi)^{2p}}\textrm{Li}_{2p+1}\left(e^{-2\pi q}\right)\;,
\end{equation}
where $B_{n}$ represent the Bernoulli numbers and $\textrm{Li}_{s}(w)$ is the polylogarithmic function. By using (\ref{31}) and (\ref{32}) in (\ref{30}) one obtains,
after a long but straightforward calculation,
\begin{equation}\label{33}
 \Re[{\cal A}_{2p}(E)]=\frac{(-1)^{p}(2p)!}{2^{2p+1}(2\pi)^{2p}}\left[\textrm{Li}_{2p+1}\left(e^{-2\pi \frac{m^{2}}{E}}\right)-2^{2p}\textrm{Li}_{2p+1}\left(e^{-\pi \frac{m^{2}}{E}}\right)\right]\;,
\end{equation}
which can finally be rewritten, according to the relation \cite{lewin}, valid for $s\in\mathbb{C}$ and $q\in\mathbb{R}$,
\begin{equation}\label{34}
  \textrm{Li}_{s}\left(e^{-2\pi q}\right)=2^{s-1}\left[\textrm{Li}_{s}\left(e^{-\pi q}\right)+\textrm{Li}_{s}\left(-e^{-\pi q}\right)\right]\;,
\end{equation}
simply as
\begin{equation}\label{35}
  \Re[{\cal A}_{2p}(E)]=\frac{(-1)^{p}(2p)!}{2(2\pi)^{2p}}\textrm{Li}_{2p+1}\left(-e^{-\pi \frac{m^{2}}{E}}\right)\;.
\end{equation}

Now for $(l+1-k)=2p+1$, $p\in\mathbb{N}_{0}$, we can write
\begin{eqnarray}\label{36}
&&\Im\Bigg[(-2iE)^{2p+1}\Bigg\{
 \left[H_{2p}-\ln\left(-i\frac{2E}{\mu^{2}}\right)\right]\zeta_{H}\left(-2p\,;\frac{1}{2}+i\frac{m^{2}}{2E}\right)+\zeta'_{H}\left(-2p\,;\frac{1}{2}+i\frac{m^{2}}{2E}\right)\Bigg\}\Bigg]\nonumber\\
 &=&(-1)^{p+1}(2E)^{2p+2}\Im[{\cal A}_{2p+1}(E)]\;.
\end{eqnarray}
By utilizing, once more, the relation (\ref{29a}) one has
\begin{eqnarray}\label{37}
\Im[{\cal A}_{2p+1}(E)]&=&\left[2^{-2p-1}\Im\zeta_{H}\left(-2p-1\,;i\frac{m^{2}}{E}\right)-\Im\zeta_{H}\left(-2p-1\,;i\frac{m^{2}}{2E}\right)\right]\left[H_{2p+1}-\ln\left(\frac{2E}{\mu^{2}}\right)\right]\nonumber\\
&-&\frac{\pi}{2}\left[2^{-2p-1}\Re\zeta_{H}\left(-2p-1\,;i\frac{m^{2}}{E}\right)-\Re\zeta_{H}\left(-2p-1\,;i\frac{m^{2}}{2E}\right)\right]+2^{-2p-1}\ln 2\Im\zeta_{H}\left(-2p-1\,;i\frac{m^{2}}{E}\right)\nonumber\\
&+&2^{-2p-1}\Im\zeta'_{H}\left(-2p-1\,;i\frac{m^{2}}{E}\right)-\Im\zeta'_{H}\left(-2p-1\,;i\frac{m^{2}}{2E}\right)\;.
\end{eqnarray}
The real and imaginary parts of the Hurwitz zeta function that appear in (\ref{37}) can be explicitly evaluated and read \cite{fucci11a}
\begin{equation}\label{38}
  \Im\zeta_{H}(-2p-1\,; iq)=\frac{(-1)^{p}}{2}q^{2p+1}\;,
  \end{equation}
  \begin{equation}
  \Re\zeta_{H}(-2p-1\,; iq)=(2p+1)!\sum_{j=0}^{p}\frac{(-1)^{j+1}B_{2(p-j+1)}}{[2(p-j+1)]!(2j)!}q^{2j}+\frac{(-1)^{p}}{2(p+1)}q^{2p+2}\;,
\end{equation}
and
\begin{equation}\label{39}
  \Im\zeta'_{H}(-2p-1\,; iq)=-\frac{\pi}{2}\Re\zeta_{H}(-2p-1\,; iq)-\frac{(-1)^{p}}{2}q^{2p+1}\ln q+(-1)^{p}\frac{(2p+1)!}{2(2\pi)^{2p+1}}\textrm{Li}_{2p+2}\left(e^{-2\pi q}\right)\;.
\end{equation}
The substitution of (\ref{38}) through (\ref{39}) in the expression (\ref{37}) together with the result (\ref{34}) leads to the formula, valid when $(l+1-k)=2p+1$,
\begin{equation}\label{40}
  \Im[{\cal A}_{2p+1}(E)]=\frac{(-1)^{p}(2p+1)!}{2(2\pi)^{2p+1}}\textrm{Li}_{2p+2}\left(-e^{-\pi \frac{m^{2}}{E}}\right)\;.
\end{equation}
By using (\ref{35}) and (\ref{40}) in (\ref{26}) we can finally write the imaginary part of the first term of (\ref{24}) as
\begin{eqnarray}\label{41}
\lefteqn{\Im\Bigg[\frac{2(-2iE)^{l+2}}{(4\pi)^{l+2}}\sum_{k=0}^{l+1}a_{k}^{N}\frac{(-1)^{l+1-k}(-2iE)^{-k}}{(l+1-k)!}\Bigg\{
 \left[H_{l+1-k}-\ln\left(-i\frac{2E}{\mu^{2}}\right)\right]\zeta_{H}\left(k-l-1;\frac{1}{2}+i\frac{m^{2}}{2E}\right)}\nonumber\\
 &+&\zeta'_{H}\left(k-l-1;\frac{1}{2}+i\frac{m^{2}}{2E}\right)\Bigg\}\Bigg]=-\frac{E^{l+2}}{(2\pi)^{2l+3}}\sum_{k=0}^{l+1}a_{k}^{N}\left(\frac{E}{\pi}\right)^{-k}\textrm{Li}_{l+2-k}\left(-e^{-\pi \frac{m^{2}}{E}}\right)\;.
\end{eqnarray}

We can now focus our attention to the second term of (\ref{24}). The formula (\ref{29a}) allows us to write, for the finite part of the Hurwitz zeta function at $s=1$,
\begin{equation}\label{42}
  \textrm{FP}\,\zeta_{H}\left(1;\frac{1}{2}+i\frac{m^{2}}{2E}\right)=2\ln 2+2\textrm{FP}\,\zeta_{H}\left(1;i\frac{m^{2}}{E}\right)-\textrm{FP}\,\zeta_{H}\left(1;i\frac{m^{2}}{2E}\right)\;.
\end{equation}
To evaluate the finite parts appearing on the right-hand side of the previous equation, we first notice that the Hurwitz zeta function of imaginary second argument
can be expressed, for $s\in\mathbb{C}$ and $q\in(0,1)$, as follows \cite{fucci11a}
\begin{equation}\label{43}
  \zeta_{H}(s;iq)=\frac{\Gamma(1-s)}{(2\pi)^{1-s}}\left[\sin\left(\frac{\pi s}{2}\right)F(s,q)+i\cos\left(\frac{\pi s}{2}\right)G(s,q)\right]\;,
\end{equation}
where
\begin{equation}\label{44}
  F(s,q)=\textrm{Li}_{1-s}\left(e^{2\pi q}\right)+\textrm{Li}_{1-s}\left(e^{-2\pi q}\right)\;,\quad \textrm{and}\quad
  G(s,q)=\textrm{Li}_{1-s}\left(e^{2\pi q}\right)-\textrm{Li}_{1-s}\left(e^{-2\pi q}\right)\;.
\end{equation}
By setting $s=1-\varepsilon$ in (\ref{43}), one obtains, for $\varepsilon\to 0$,
\begin{equation}\label{45}
  \zeta_{H}(1-\varepsilon;iq)=-\frac{1}{\varepsilon}+\gamma+\log 2\pi-F'(1,q)+i\frac{\pi}{2}G(1,q)+O(\varepsilon)\;,
\end{equation}
where $\gamma$ is the Euler-Mascheroni constant and it can be proved \cite{fucci11a} that
\begin{equation}\label{46}
  F'(1,q)=\frac{i}{2q}-2\sum_{k=0}^{\infty}\frac{\zeta'(-2k)}{(2k)!}(2\pi q)^{k}\;, \quad\textrm{and}\quad G(1,q)=-\frac{1}{\tanh(\pi q)}\;.
\end{equation}
 By using the result (\ref{45}) in (\ref{42}) it is not very difficult to obtain
\begin{equation}\label{47}
   \textrm{FP}\,\zeta_{H}\left(1;\frac{1}{2}+i\frac{m^{2}}{2E}\right)=\ln 8\pi +\gamma+2\sum_{k=0}^{\infty}\frac{\zeta'(-2k)}{(2k)!}(2^{2k+1}-1)\left(\frac{\pi m^{2}}{E}\right)^{k}
   -\frac{i\pi}{2}\tanh\left(\frac{\pi m^{2}}{2E}\right)\;,
\end{equation}
from which one can easily extract the needed imaginary part. We can, therefore, conclude that the imaginary part of the second term of (\ref{24}) reads
\begin{equation}\label{48}
  \Im\left\{\frac{2}{(4\pi)^{l+2}}a_{l+2}^{N}\left[\textrm{FP}\,\zeta_{H}\left(1;\frac{1}{2}+i\frac{m^{2}}{2E}\right)-\ln\left(-i\frac{2E}{\mu^{2}}\right)\right]\right\}
  =\frac{1}{2(4\pi)^{l+1}}a_{l+2}^{N}\textrm{Li}_{0}\left(-e^{-\pi\frac{m^{2}}{E}}\right)\;,
\end{equation}
which can be obtained by noticing that
\begin{equation}
\frac{1}{2}\left[1-\tanh\left(\frac{\pi m^{2}}{2E}\right)\right]=\textrm{Li}_{0}\left(-e^{-\pi\frac{m^{2}}{E}}\right)\;.
\end{equation}

Let us continue the discussion with the evaluation of the imaginary part of the complex-valued element in the third term of (\ref{24}).
We use the general formula (\ref{25a}) to write
\begin{eqnarray}\label{49}
\lefteqn{\Im\Bigg[(-2iE)^{l-k+\frac{3}{2}}\zeta_{H}\left(k-l-\frac{1}{2};\frac{1}{2}+i\frac{m^{2}}{2E}\right)\Bigg]}\nonumber\\
&=&(2E)^{l-k+\frac{3}{2}}\Bigg\{\sin\left[\frac{\pi}{2}\left(k-l-\frac{1}{2}\right)\right]\left[2^{k-l-\frac{1}{2}}\Im\zeta_{H}\left(k-l-\frac{1}{2};i\frac{m^{2}}{E}\right)-\Im\zeta_{H}\left(k-l-\frac{1}{2};i\frac{m^{2}}{2E}\right)\right]\nonumber\\
&-&\cos\left[\frac{\pi}{2}\left(k-l-\frac{1}{2}\right)\right]\left[2^{k-l-\frac{1}{2}}\Re\zeta_{H}\left(k-l-\frac{1}{2};i\frac{m^{2}}{E}\right)-\Re\zeta_{H}\left(k-l-\frac{1}{2};i\frac{m^{2}}{2E}\right)\right]\Bigg\}\;.
\end{eqnarray}
As it is clear from (\ref{43}) the Hurwitz zeta functions in (\ref{49}) can be written in terms of the polylogarithmic functions. We would like to point out that
for $s\in\mathbb{R}$ and for $q\in(0,1)$, the function $\textrm{Li}_{1-s}(e^{-2\pi q})$ is real while $\textrm{Li}_{1-s}(e^{2\pi q})$ becomes a complex function for
$s<1$ and a real function for $s\geq 1$. Hence, for $s<1$ one finds \cite{fucci11a}
\begin{eqnarray}
  \Re\zeta_{H}(s,iq)&=&\frac{\Gamma(1-s)}{(2\pi)^{1-s}}\left[\sin\left(\frac{\pi}{2}s\right)\Re F(s,q)-\cos\left(\frac{\pi}{2}s\right)\Im G(s,q)\right]\;,\label{50}\\
   \Im\zeta_{H}(s,iq)&=&\frac{\Gamma(1-s)}{(2\pi)^{1-s}}\left[\cos\left(\frac{\pi}{2}s\right)\Re G(s,q)+\sin\left(\frac{\pi}{2}s\right)\Im F(s,q)\right]\;,\label{50a}
\end{eqnarray}
where
\begin{eqnarray}
  \Re F(s,q)&=&\frac{\pi}{\tan(\pi s)}\frac{(2\pi q)^{-s}}{\Gamma(1-s)}+\sum_{k=0}^{\infty}\frac{\zeta_{R}(1-s-k)}{k!}(2\pi q)^{k}+\textrm{Li}_{1-s}\left(e^{-2\pi q}\right)\;,\label{51}\\
  \Im F(s,q)&=&-\pi\frac{(2\pi q)^{-s}}{\Gamma(1-s)}\;,\label{51a}
  \end{eqnarray}
and
\begin{equation}\label{52}
 \Re G(s,q)=\Re F(s,q)-2\textrm{Li}_{1-s}\left(e^{-2\pi q}\right)\;,\quad \Im G(s,q)=\Im F(s,q)\;.
\end{equation}
When $s\geq 1$, $\textrm{Li}_{1-s}(e^{2\pi q})$ is a real function when $q\in(0,1)$ and the expression (\ref{43}) is sufficient to readily
find the real and imaginary parts.
These results allow us to proceed with the explicit calculation of the quantity in (\ref{49}). When $k-l-1/2<1$ we obtain
\begin{eqnarray}\label{53}
\lefteqn{\sin\left[\frac{\pi}{2}\left(k-l-\frac{1}{2}\right)\right]\left[2^{k-l-\frac{1}{2}}\Im\zeta_{H}\left(k-l-\frac{1}{2};i\frac{m^{2}}{E}\right)-\Im\zeta_{H}\left(k-l-\frac{1}{2};i\frac{m^{2}}{2E}\right)\right]}\nonumber\\
&=&\frac{\Gamma\left(l+\frac{3}{2}-k\right)}{2(2\pi)^{l+\frac{3}{2}-k}}\Bigg\{(-1)^{l-k+1}\left[2^{k-l-\frac{1}{2}}\Re G\left(k-l-\frac{1}{2};\frac{m^{2}}{E}\right)
-\Re G\left(k-l-\frac{1}{2};\frac{m^{2}}{2E}\right)\right]\nonumber\\
&+&2^{k-l-\frac{1}{2}}\Im F\left(k-l-\frac{1}{2};\frac{m^{2}}{E}\right)
-\Im F\left(k-l-\frac{1}{2};\frac{m^{2}}{2E}\right)\Bigg\}\;.
\end{eqnarray}
By exploiting the results (\ref{51}) through (\ref{52}) and also (\ref{34}) it is not very difficult to get
\begin{eqnarray}\label{54}
\lefteqn{\sin\left[\frac{\pi}{2}\left(k-l-\frac{1}{2}\right)\right]\left[2^{k-l-\frac{1}{2}}\Im\zeta_{H}\left(k-l-\frac{1}{2};i\frac{m^{2}}{E}\right)-\Im\zeta_{H}\left(k-l-\frac{1}{2};i\frac{m^{2}}{2E}\right)\right]}\nonumber\\
&=&\frac{(-1)^{l-k+1}\Gamma\left(l+\frac{3}{2}-k\right)}{2(2\pi)^{l+\frac{3}{2}-k}}\left[\sum_{j=0}^{\infty}\frac{\zeta\left(\frac{3}{2}+l-k-j\right)}{j!}
\left(\frac{\pi m^{2}}{E}\right)^{j}\left(2^{j-l+k-\frac{1}{2}}-1\right)-\textrm{Li}_{\frac{3}{2}+l-k}\left(-e^{-\pi\frac{m^{2}}{E}}\right)\right]\;.
\end{eqnarray}

Similarly, we can write
\begin{eqnarray}\label{55}
\lefteqn{\cos\left[\frac{\pi}{2}\left(k-l-\frac{1}{2}\right)\right]\left[2^{k-l-\frac{1}{2}}\Re\zeta_{H}\left(k-l-\frac{1}{2};i\frac{m^{2}}{E}\right)-\Re\zeta_{H}\left(k-l-\frac{1}{2};i\frac{m^{2}}{2E}\right)\right]}\nonumber\\
&=&\frac{\Gamma\left(l+\frac{3}{2}-k\right)}{2(2\pi)^{l+\frac{3}{2}-k}}\Bigg\{-2^{k-l-\frac{1}{2}}\Im G\left(k-l-\frac{1}{2};\frac{m^{2}}{E}\right)
+\Im G\left(k-l-\frac{1}{2};\frac{m^{2}}{2E}\right)\nonumber\\
&+&(-1)^{l-k+1}\left[2^{k-l-\frac{1}{2}}\Re F\left(k-l-\frac{1}{2};\frac{m^{2}}{E}\right)
-\Re F\left(k-l-\frac{1}{2};\frac{m^{2}}{2E}\right)\right]\Bigg\}\;.
\end{eqnarray}
By using, once again (\ref{34}) and (\ref{51}) through (\ref{52}) we arrive at the result
\begin{eqnarray}\label{56}
\lefteqn{\cos\left[\frac{\pi}{2}\left(k-l-\frac{1}{2}\right)\right]\left[2^{k-l-\frac{1}{2}}\Re\zeta_{H}\left(k-l-\frac{1}{2};i\frac{m^{2}}{E}\right)-\Re\zeta_{H}\left(k-l-\frac{1}{2};i\frac{m^{2}}{2E}\right)\right]}\nonumber\\
&=&\frac{(-1)^{l-k+1}\Gamma\left(l+\frac{3}{2}-k\right)}{2(2\pi)^{l+\frac{3}{2}-k}}\left[\sum_{j=0}^{\infty}\frac{\zeta\left(\frac{3}{2}+l-k-j\right)}{j!}
\left(\frac{\pi m^{2}}{E}\right)^{j}\left(2^{j-l+k-\frac{1}{2}}-1\right)+\textrm{Li}_{\frac{3}{2}+l-k}\left(-e^{-\pi\frac{m^{2}}{E}}\right)\right]\;.
\end{eqnarray}
By substituting (\ref{54}) and (\ref{56}) in (\ref{49}) we obtain, when $k-l-1/2<1$, the expression
\begin{equation}\label{57}
\Im\Bigg[(-2iE)^{l-k+\frac{3}{2}}\zeta_{H}\left(k-l-\frac{1}{2};\frac{1}{2}+i\frac{m^{2}}{2E}\right)\Bigg]
=(2E)^{l-k+\frac{3}{2}}\frac{(-1)^{l-k}\Gamma\left(l+\frac{3}{2}-k\right)}{(2\pi)^{l+\frac{3}{2}-k}}\textrm{Li}_{\frac{3}{2}+l-k}\left(-e^{-\pi\frac{m^{2}}{E}}\right)\;.
\end{equation}

We consider, now, the case $k-l-1/2>1$. In this situation the real and imaginary parts of the Hurwitz zeta function of imaginary second argument
can be extracted directly from (\ref{43}). In more details, one has
\begin{eqnarray}\label{58}
\lefteqn{\Im\Bigg[(-2iE)^{l-k+\frac{3}{2}}\zeta_{H}\left(k-l-\frac{1}{2};\frac{1}{2}+i\frac{m^{2}}{2E}\right)\Bigg]}\nonumber\\
&=&(2E)^{l-k+\frac{3}{2}}\frac{(-1)^{l-k+1}\Gamma\left(l+\frac{3}{2}-k\right)}{2(2\pi)^{l+\frac{3}{2}-k}}\Bigg[2^{k-l-\frac{1}{2}}G\left(k-l-\frac{1}{2};\frac{m^{2}}{E}\right)
-G\left(k-l-\frac{1}{2};\frac{m^{2}}{2E}\right)\nonumber\\
&-&2^{k-l-\frac{1}{2}}F\left(k-l-\frac{1}{2};\frac{m^{2}}{E}\right)+F\left(k-l-\frac{1}{2};\frac{m^{2}}{2E}\right)\Bigg]\;.
\end{eqnarray}
According to (\ref{44}) and (\ref{34})
\begin{equation}\label{59}
2^{k-l-\frac{1}{2}}G\left(k-l-\frac{1}{2};\frac{m^{2}}{E}\right)
-G\left(k-l-\frac{1}{2};\frac{m^{2}}{2E}\right)=\textrm{Li}_{\frac{3}{2}-k+l}\left(-e^{\pi\frac{m^{2}}{E}}\right)-\textrm{Li}_{\frac{3}{2}-k+l}\left(-e^{-\pi\frac{m^{2}}{E}}\right)\;,
\end{equation}
and
\begin{equation}\label{60}
2^{k-l-\frac{1}{2}}F\left(k-l-\frac{1}{2};\frac{m^{2}}{E}\right)
-F\left(k-l-\frac{1}{2};\frac{m^{2}}{2E}\right)=\textrm{Li}_{\frac{3}{2}-k+l}\left(-e^{\pi\frac{m^{2}}{E}}\right)+\textrm{Li}_{\frac{3}{2}-k+l}\left(-e^{-\pi\frac{m^{2}}{E}}\right)\;.
\end{equation}
The last two results allows us to write, also for $k-l-1/2>1$,
\begin{eqnarray}\label{61}
\Im\Bigg[(-2iE)^{l-k+\frac{3}{2}}\zeta_{H}\left(k-l-\frac{1}{2};\frac{1}{2}+i\frac{m^{2}}{2E}\right)\Bigg]=(2E)^{l-k+\frac{3}{2}}\frac{(-1)^{l-k}\Gamma\left(l+\frac{3}{2}-k\right)}{(2\pi)^{l+\frac{3}{2}-k}}\textrm{Li}_{\frac{3}{2}+l-k}\left(-e^{-\pi\frac{m^{2}}{E}}\right)\;.
\end{eqnarray}
Hence, from (\ref{57}) and (\ref{61}) we can express the imaginary part of the third term in (\ref{24}) as
\begin{eqnarray}\label{62}
&&\Im\left\{\frac{2(-2iE)^{l+2}}{(4\pi)^{l+2}}\sum_{k=0}^{\infty}a_{k+\frac{1}{2}}^{N}(-2iE)^{-k-\frac{1}{2}}
 \Gamma\left(k-l-\frac{1}{2}\right)\zeta_{H}\left(k-l-\frac{1}{2};\frac{1}{2}+i\frac{m^{2}}{2E}\right)\right\}\nonumber\\
 &=&-\frac{E^{l+\frac{3}{2}}}{2\sqrt{\pi}(2\pi)^{2l+2}}\sum_{k=0}^{\infty}a_{k+\frac{1}{2}}^{N}\left(\frac{\pi}{E}\right)^{k}\textrm{Li}_{\frac{3}{2}-k+l}\left(-e^{-\pi\frac{m^{2}}{E}}\right)\;.
\end{eqnarray}

Lastly, we evaluate the imaginary part of the complex-valued component of the fourth term in (\ref{24}).
In particular we find
\begin{eqnarray}\label{63}
\Im\left\{(-2iE)^{-k+l+2}\zeta_{H}\left(k-l-1;\frac{1}{2}+i\frac{m^{2}}{2E}\right)\right\}
&=&(2E)^{-k+l+2}\Bigg\{\sin\left[\frac{\pi}{2}(k-l-1)\right]\Im\zeta_{H}\left(k-l-1;\frac{1}{2}+i\frac{m^{2}}{2E}\right)\nonumber\\
&-&\cos\left[\frac{\pi}{2}(k-l-1)\right]\Re\zeta_{H}\left(k-l-1;\frac{1}{2}+i\frac{m^{2}}{2E}\right)\Bigg\}\;.
\end{eqnarray}
Once again, It is convenient to separate even and odd values of the quantity $k-l-1$. When $k-l-1=2p$, $p\in\mathbb{N}_{0}$ we have, by using (\ref{29a}),
\begin{eqnarray}\label{64}
\Im\left\{(-2iE)^{-2p+1}\zeta_{H}\left(2p\,;\frac{1}{2}+i\frac{m^{2}}{2E}\right)\right\}
&=&(2E)^{-2p+1}(-1)^{p+1}\left[2^{2p}\Re\zeta_{H}\left(2p;i\frac{m^{2}}{E}\right)-\Re\zeta_{H}\left(2p;i\frac{m^{2}}{2E}\right)\right]\;,\;\;\;\;\;\;
\end{eqnarray}
while for $k-l-1=2p+1$, $p\in\mathbb{N}_{0}$, we obtain instead
\begin{eqnarray}\label{65}
\lefteqn{\Im\left\{(-2iE)^{-2p}\zeta_{H}\left(2p+1\,;\frac{1}{2}+i\frac{m^{2}}{2E}\right)\right\}}\nonumber\\
&=&(2E)^{-2p}(-1)^{p}\left[2^{2p+1}\Im\zeta_{H}\left(2p+1;i\frac{m^{2}}{E}\right)-\Im\zeta_{H}\left(2p+1;i\frac{m^{2}}{2E}\right)\right]\;.
\end{eqnarray}
For positive integers and for $q\in(0,1)$, one can prove that \cite{fucci11a}
\begin{equation}\label{66}
  \Re\zeta_{H}(2p,iq)=(-1)^{p}\frac{(2\pi)^{2p}}{4(2p-1)!}F(2p,q)+\frac{(-1)^{p}}{2}q^{-2p}\;,
\end{equation}
and
\begin{equation}\label{67}
 \Im\zeta_{H}(2p+1,iq)=(-1)^{p}\frac{(2\pi)^{2p+1}}{4(2p)!}G(2p+1,q)+\frac{(-1)^{p+1}}{2}x^{-2p-1}\;.
\end{equation}
From the definitions (\ref{44}) and the relation (\ref{34}) one finds that
\begin{equation}\label{68}
2^{2p}\Re\zeta_{H}\left(2p;i\frac{m^{2}}{E}\right)-\Re\zeta_{H}\left(2p;i\frac{m^{2}}{2E}\right)=
\frac{(-1)^{p}(2\pi)^{2p}}{4(2p-1)!}\left[\textrm{Li}_{1-2p}\left(-e^{\pi\frac{m^{2}}{E}}\right)+\textrm{Li}_{1-2p}\left(-e^{-\pi\frac{m^{2}}{E}}\right)\right]\;,
\end{equation}
and
\begin{equation}\label{69}
2^{2p+1}\Im\zeta_{H}\left(2p+1;i\frac{m^{2}}{E}\right)-\Im\zeta_{H}\left(2p+1;i\frac{m^{2}}{2E}\right)=
\frac{(-1)^{p}(2\pi)^{2p+1}}{4(2p)!}\left[\textrm{Li}_{-2p}\left(-e^{\pi\frac{m^{2}}{E}}\right)-\textrm{Li}_{-2p}\left(-e^{-\pi\frac{m^{2}}{E}}\right)\right]\;.
\end{equation}
By using the last results in (\ref{64}) and (\ref{65}) it is not very difficult to get
\begin{eqnarray}\label{70}
\lefteqn{\Im\Bigg\{\frac{2(-2iE)^{l+2}}{(4\pi)^{l+2}}\sum_{k=l+3}^{\infty}a_{k}^{N}(-2iE)^{-k}
 \Gamma\left(k-l-1\right)\zeta_{H}\left(k-l-1;\frac{1}{2}+i\frac{m^{2}}{2E}\right)\Bigg\}}\nonumber\\
 &=&\frac{\pi}{(4\pi)^{l+2}}\sum_{k=l+3}^{\infty}\left(\frac{\pi}{E}\right)^{k-l-2}\left[(-1)^{k-l-2}\textrm{Li}_{l-k+2}\left(-e^{\pi\frac{m^{2}}{E}}\right)-\textrm{Li}_{l-k+2}\left(-e^{-\pi\frac{m^{2}}{E}}\right)\right]\;.
\end{eqnarray}
By noticing that for $p\in\mathbb{N}^{+}$ and $z\in\mathbb{C}/\{0\}$, the polylogarithmic functions satisfy \cite{lewin}
\begin{equation}
\textrm{Li}_{-p}(z)+(-1)^{p}\textrm{Li}_{-p}\left(\frac{1}{z}\right)=0\;,
\end{equation}
the formula in (\ref{70}) can be further simplified to
\begin{eqnarray}\label{71}
&&\Im\Bigg\{\frac{2(-2iE)^{l+2}}{(4\pi)^{l+2}}\sum_{k=l+3}^{\infty}a_{k}^{N}(-2iE)^{-k}
 \Gamma\left(k-l-1\right)\zeta_{H}\left(k-l-1;\frac{1}{2}+i\frac{m^{2}}{2E}\right)\Bigg\}\nonumber\\
 &=&-\frac{2\pi}{(4\pi)^{l+2}}\sum_{k=l+3}^{\infty}a_{k}^{N}\left(\frac{\pi}{E}\right)^{k-l-2}\textrm{Li}_{l-k+2}\left(-e^{-\pi\frac{m^{2}}{E}}\right)\;.
\end{eqnarray}

We can finally collect the results in (\ref{41}), (\ref{48}), (\ref{62}), and (\ref{71}) to write the imaginary part of (\ref{24}) valid
for even dimensions $d=2l$,
\begin{eqnarray}\label{72}
&&\frac{1}{\Omega}\Im\left[\frac{\diff}{\diff s}\left[\mu^{-2s}\zeta(s)\right]\Big|_{s=0}\right]\sim-\frac{E^{l+2}}{(2\pi)^{2l+3}}\sum_{k=0}^{l+1}a_{k}^{N}\left(\frac{\pi}{E}\right)^{k}\textrm{Li}_{l+2-k}\left(-e^{-\pi \frac{m^{2}}{E}}\right)+\frac{1}{2(4\pi)^{l+1}}a_{l+2}^{N}\textrm{Li}_{0}\left(-e^{-\pi\frac{m^{2}}{E}}\right)\nonumber\\
&-&\frac{E^{l+\frac{3}{2}}}{2\sqrt{\pi}(2\pi)^{2l+2}}\sum_{k=0}^{\infty}a_{k+\frac{1}{2}}^{N}\left(\frac{\pi}{E}\right)^{k}\textrm{Li}_{l+\frac{3}{2}-k}\left(-e^{-\pi\frac{m^{2}}{E}}\right)-\frac{E^{l+2}}{2(4\pi)^{l+1}}\sum_{k=l+3}^{\infty}a_{k}^{N}\left(\frac{\pi}{E}\right)^{k}\textrm{Li}_{l+2-k}\left(-e^{-\pi\frac{m^{2}}{E}}\right)\;.
\end{eqnarray}

A similar calculation can be performed in order to obtain an expression for the imaginary part of (\ref{25}) corresponding
to the case of an odd-dimensional manifold $N$. Therefore, for the sake of brevity, we will not repeat the explicit calculations and present
the final result as
\begin{eqnarray}\label{73}
&&\frac{1}{\Omega}\Im\left[\frac{\diff}{\diff s}\left[\mu^{-2s}\zeta(s)\right]\Big|_{s=0}\right]\sim-\frac{E^{l+2}}{2\sqrt{\pi}(2\pi)^{2l+3}}\sum_{k=0}^{l+1}a_{k+\frac{1}{2}}^{N}\left(\frac{\pi}{E}\right)^{k}\textrm{Li}_{l+2-k}\left(-e^{-\pi \frac{m^{2}}{E}}\right)+\frac{1}{2(4\pi)^{l+\frac{3}{2}}}a_{l+\frac{5}{2}}^{N}\textrm{Li}_{0}\left(-e^{-\pi\frac{m^{2}}{E}}\right)\nonumber\\
&-&\frac{E^{l+\frac{5}{2}}}{(2\pi)^{2l+4}}\sum_{k=0}^{\infty}a_{k}^{N}\left(\frac{\pi}{E}\right)^{k}\textrm{Li}_{l+\frac{5}{2}-k}\left(-e^{-\pi\frac{m^{2}}{E}}\right)
-\frac{E^{l+2}}{2\sqrt{\pi}(2\pi)^{2l+3}}\sum_{k=l+3}^{\infty}a_{k+\frac{1}{2}}^{N}\left(\frac{\pi}{E}\right)^{k}\textrm{Li}_{l+2-k}\left(-e^{-\pi\frac{m^{2}}{E}}\right)\;.
\end{eqnarray}
valid when $d=2l+1$, with $l\in\mathbb{N}_{0}$.

\section{The rate of particle creation for specific manifolds $N$}\label{Sec4}

The results obtained in the last section are sufficient for discussing the rate of particle creation on product manifolds of the type $M_{4}\times N$.
While the imaginary part of the derivative of the spectral zeta function at $s=0$ is more precisely related to the vacuum persistence probability,
the rate of particle creation per unit volume is given by (\ref{72}) and (\ref{73}) with the polylogarithmic functions replaced by the first term of their defining series \cite{cohen,dunne09,holst,niki}
\begin{equation}
\textrm{Li}_{s}\left(-e^{-\pi\frac{m^{2}}{E}}\right)\approx -e^{-\pi\frac{m^{2}}{E}}\;.
\end{equation}
By recalling that $a_{0}^{N}=1$, the first few leading terms contributing to the rate of particle creation when the dimension of the
manifold $N$ is $d$, are, from (\ref{72}) and (\ref{73}),
\begin{equation}\label{74}
R\simeq \frac{E^{\frac{d}{2}+2}}{(2\pi)^{d+3}}e^{-\pi \frac{m^{2}}{E}}\left[1+\sqrt{\frac{\pi}{E}}a_{\frac{1}{2}}^{N}+\frac{\pi}{E}a_{1}^{N}
+\left(\frac{\pi}{E}\right)^{\frac{3}{2}}a_{\frac{3}{2}}^{N}+O(E^{-2})\right]\;.
\end{equation}
The above result clearly shows that the leading term describes the rate of particle creation per unit space-time volume in a flat space of dimension
$D=d+4$. The remaining subleading terms represent the corrections to the flat space rate due to the geometric characteristics of the manifold $N$.
We would like to make a comment at this point. It is well-known that the rate of particle creation in four-dimensional Minkowski space is proportional to $E^{2}$
\cite{dunne09,schwing51}. From the result presented in (\ref{74}) one can easily realize that the rate of particle creation
in the presence of $d$ extra dimensions is, instead, proportional to $E^{\frac{d}{2}+2}$. This implies that the mere presence of $d$ extra dimensions, irrespective of
their particular geometry, increases the rate of particle production by a factor of $E^{d/2}$.

The results we have obtained so far are valid for any smooth, compact Riemmanina manifold $N$ and can, therefore, be used to obtain
explicit formulas for the rate of particle creation for specific manifolds. Below we illustrate a few simple cases.
\begin{paragraph}{Flat manifold $N$.}
If we assume that the manifold $N$, representing the extra dimensions, is flat, then all the coefficients $a_{k/2}^{N}$ of the small-$t$ expansion
of the trace of the heat kernel vanish identically \cite{kirsten01} except for the first one for which the relation $a_{0}^{N}=1$ holds.
In this case the formula in (\ref{72}) provides a very simple expression for the rate of particle creation. In more details, for any dimension $d$ we have
\begin{equation}\label{76}
R\simeq\frac{E^{\frac{d}{2}+2}}{(2\pi)^{d+3}}e^{-\pi\frac{m^{2}}{E}}\;,
\end{equation}
which, as one should expect, coincides with the rate of particle creation in a $(d+4)$-dimensional Minkowski space \cite{gavrilov95}.
\end{paragraph}
\begin{paragraph}{Compact manifold $N$ without boundary.}
If the manifold $N$ is considered to be compact and without boundary then all the coefficients $a_{k+\frac{1}{2}}^{N}$, with $k\in\mathbb{N}_{0}$ vanish
since they incorporate only the geometric information of the boundary of $N$ \cite{kirsten01}.
In this situation only the coefficients $a_{k}^{N}$ with positive integer index are present in the expansions (\ref{72}) and (\ref{73}) which provide the following
expression for the rate of particle creation
\begin{equation}\label{77}
R\simeq\frac{E^{\frac{d}{2}+2}}{(2\pi)^{d+3}}e^{-\pi\frac{m^{2}}{E}}\left[1+\frac{\pi}{E}a_{1}^{N}+\left(\frac{\pi}{E}\right)^{2}a_{2}^{N}+O(E^{-3})\right]\;.
\end{equation}
The coefficient $a_{1}^{N}$ of the asymptotic expansion of the heat kernel is proportional
to the integral of the scalar curvature of $N$ \cite{gilkey95,kirsten01}. From (\ref{77}) it is not difficult to realize that if $N$ has positive scalar curvature, then
the rate of particle creation is enhanced compared to the rate in a $D$-dimensional flat Minkowski space. The rate is, instead, decreased if
the manifold $N$ has negative scalar curvature. We would like to point out here, that the result obtained for the rate of particle creation in (\ref{74})
does not hold for strongly curved manifolds, that is for manifolds for which $R\sim E$. In fact, the coefficients $a_{j}^{N}$ are proportional to integrals
of geometric quantities constructed from the $j$-th power of the curvature of $N$. It is clear from the previous formulas that if $R\sim E$, then every term in (\ref{74})
would have the same order of magnitude of the term preceding it and, hence, the entire series would have to be taken into account.
\end{paragraph}
\begin{paragraph}{Manifold $N$ as a $d$-dimensional ball.}
We assume, now, that the manifold $N$ is a $d$-dimensional ball of radius $R$, $B^{d}=\{x\in\mathbb{R}^{d}; |x|\leq R\}$.
For this case, the heat kernel coefficients $A^{B^{d}}_{k/2}$ are well-known \cite{bordag96} for a variety of boundary conditions.
To obtain $a_{k/2}^{B^{d}}$ from $A^{B^{d}}_{k/2}$ in \cite{bordag96} we need to recall that for a $d$-dimensional ball of radius $R$,
$\textrm{Vol}\left(B^{d}\right)=\pi^{d/2}R^{d}/\Gamma(d/2+1)$ and that $a_{k/2}^{B^{d}}=A^{B^{d}}_{k/2}/\textrm{Vol}\left(B^{d}\right)$.
For Dirichlet boundary conditions and $d=3$ the rate of particle creation can be found to be
\begin{equation}\label{77a}
R\simeq \frac{E^{\frac{7}{2}}}{(2\pi)^{6}}e^{-\pi \frac{m^{2}}{E}}\left[1-\frac{3\pi}{2R E^{1/2}}+\frac{2\pi}{R^{2} E}
-\frac{\pi^{2}}{8 R^{3} E^{3/2}}-\frac{4\pi^{2}}{105 R^{4} E^{2}}-\frac{\pi^{3}}{160 R^{5} E^{5/2}}+O(E^{-3})\right]\;.
\end{equation}
For $d=4$ and Dirichlet boundary conditions we have, instead,
\begin{equation}\label{78}
R\simeq \frac{E^{4}}{(2\pi)^{7}}e^{-\pi \frac{m^{2}}{E}}\left[1-\frac{2\pi}{R E^{1/2}}+\frac{4\pi}{R^{2} E}
-\frac{11\pi^{2}}{16 R^{3} E^{3/2}}-\frac{8\pi^{2}}{45 R^{4} E^{2}}-\frac{35\pi^{3}}{2084 R^{5} E^{5/2}}+O(E^{-3})\right]\;.
\end{equation}
When $d=5$, we obtain, for Dirichlet boundary conditions, the following particle creation rate
\begin{equation}\label{79}
R\simeq \frac{E^{\frac{9}{2}}}{(2\pi)^{8}}e^{-\pi \frac{m^{2}}{E}}\left[1-\frac{5\pi}{2R E^{1/2}}+\frac{20\pi}{3R^{2} E}
-\frac{15\pi^{2}}{8 R^{3} E^{3/2}}-\frac{16\pi^{2}}{63 R^{4} E^{2}}+\frac{17\pi^{3}}{192 R^{5} E^{5/2}}+O(E^{-3})\right]\;.
\end{equation}
By utilizing the results of \cite{bordag96} one can obtain similar rates of particle creation for Neumann and Robin boundary conditions.
It is interesting to notice that the corrections to the flat $D$-dimensional particle creation rate due to the geometry of the ball are
proportional to negative integer powers of $R E^{1/2}$ regardless of the dimension $d$ of $B^{d}$. In addition,
the results (\ref{77a})-(\ref{78}) show that the leading correction in $d=3,4,5$ is always negative which implies that, to that order,
the flat $D$-dimensional particle creation rate is diminished by the presence of the $d$-dimensional ball. We would like to point out
that this is a feature of Dirichlet boundary conditions. In fact, if we consider Neumann boundary conditions the leading correction in $d=3,4,5$
is, in that case, always positive (see \cite{bordag96}). This means that for Neumann boundary conditions the flat $D$-dimensional particle creation rate is
enhanced, to the first subleading order, by the presence of the ball.
\end{paragraph}

\section{Conclusions}

In this work we have analyzed the rate of particle creation associated with a massive scalar field propagating on a product
manifold $M_{4}\times N$ under the influence of a uniform electric field. We exploited the spectral zeta function regularization technique in which
the one-loop effective action of the system under consideration is expressed in terms of the derivative of the spectral zeta function associated with the dynamical operator of
the quantum field. To obtain a well-defined spectral problem we computed the relevant spectral zeta function in an Euclidean setting and then performed an inverse Wick rotation
back to Minkowski space. In doing so, the one-loop effective action acquires an imaginary part which indicates a violation of unitarity in the system and is related to the rate of particle creation. The evaluation and analysis of the abovementioned imaginary part was the main focus of this work.
The explicit calculations we have performed in Section \ref{Sec3} were made possible by the fact that the spectral zeta function was espressed in terms of the Hurwitz zeta function of
imaginary second argument and that formulas for the latter were already known in the literature.
The results found for the imaginary part of the one-loop effective action were directly applied to the evaluation, in Section \ref{Sec4} of the rate of particle creation for 
general manifolds $N$ and then for a few simple particular cases. In general, we have found that the rate of particle creation $R$ of a massive scalar field by a uniform electric 
field is always enhanced by the presence of the manifold $N$ representing the extra dimensions. More precisely, the leading behavior of the rate $R$ is influenced only by the 
dimension of $N$ while the subleading corrections depend on the particular geometry of the extra dimensions through the coefficients of the heat kernel asymptotic expansion.
We found that the first correction to the rate of particle creation is proportional to the $a_{1/2}^{N}$ coefficient. This suggests that the boundary conditions 
the field obeys on $\partial N$ play an important role in the ambit of particle creation in the presence of extra dimensions since, as is well-known, $a_{1/2}^{N}$ encodes geometric
information about the boundary of $N$ and the specific boundary conditions. It is interesting to mention the following: The results for the rate $R$ obtained here could be used as the 
basis for an indirect observation of extra dimensions. Let us assume that one could produce strong enough electric fields. In this case the rate of particle creation could be measured experimentally. One would expect that such measured rate would be proportional, according to Schwinger's results, to $E^2$. However, if any deviation from the $E^2$ behavior was to be observed, that would indicate the existence of extra dimensions. In addition, the number of extra dimensions would be twice the measured deviation.       

The analysis performed here could be extended to include different types of manifolds. If fact, it would be interesting to analyze the 
rate of particle creation when $N$ is a manifold possessing one or more singular points. One could specifically study the influence that 
geometric singularities have on particle creation. Another generalization to our results which would be worth exploring consists in replacing the 
product manifold with a warped product manifold $M_{4}\times_{f} N$ and a warping function $f$. Manifolds with this type of geometry are very relevant in the ambit of Randall-Sundrum 
models in string theory. It would be interesting, in this situation, to analyze the dependence of the rate of particle creation on the warping function and the manifold $N$.   
It is to be expected, however, that the analysis of the rate of particle creation for both cases suggested above would be somewhat more technically involved than
the one presented here. In fact, in the more general cases the one-loop effective action is not expected to be expressed in terms of the Hurwitz zeta function. This would make 
extracting the imaginary part needed for the rate of particle creation more difficult. 
An additional extension of our results involves the computation of the rate of particle creation for a spinor field propagating on a product manifold due to a uniform electric field.
This analysis should be rather straightforward as we expect that the spinor one-loop effective action is expressed in terms of the Hurwitz zeta function. 
Extracting the imaginary part for the computation of the rate of particle creation should follow the same arguments outlined in this paper without additional technical complications.

\begin{acknowledgments}
This research was partially funded by the ECU Research and Creative Activities Award.
\end{acknowledgments}


\begin{thebibliography}{99}

\bibitem{avramidi09} Avramidi I. G., and Fucci G., Low-energy effective action in nonperturbative electrodynamics in curved space-time,
\emph{J. Math. Phys.} {\bf 50} 102302 (2009)

\bibitem{bb} Becker K., Becker M., and Schwarz J. H., \emph{String Theory and M-Theory: A Modern Introduction}, (Cambridge University Press, Cambridge, 2007) 

\bibitem{blau91} Blau S. K., Visser M., and Wipf A., Analytic results for the effective action, \emph{Int. J. Mod. Phys. A} {\bf 30}, 5409 (1991)

\bibitem{bordag96} Bordag M., Elizalde E., and Kirsten K., Heat kernel coefficients of the Laplace operator on the $D$-dimensional ball,
\emph{J. Math. Phys.} {\bf 37}, 895 (1996)

\bibitem{brezin70}  Brezin E. and Itzykson C., Pair production in vacuum by an alternating field,
\emph{Phys. Rev.} {\bf D2}, 1191 (1970)

\bibitem{byt} Bytsenko A. A., Cognola G., Elizalde E., Moretti V., and Zerbini S., \emph{Analytic Aspects of Quantum Fields}, (World Scientific Publishing, Singapore, 2003)

\bibitem{cohen}  Cohen T. D. and McGady D. A., The Schwinger mechanism revisited,  	\emph{Phys.Rev.} {\bf D78}, 036008 (2008)

\bibitem{deff} Deffayet C., Dvali G., and Gabadadze G., Accelerated universe from gravity leaking to extra dimensions, \emph{Phys. Rev.} {\bf D65}, 044023 (2002)

\bibitem{dewitt} DeWitt B. S., \emph{ The Global Approach to Quantum Field Theory}, (Oxford University Press, Oxford, 2003)

\bibitem{dumlu10} Dumlu C. K. and Dunne G. V., The Stokes phenomenon and Schwinger vacuum pair production in time-dependent laser pulses,
\emph{Phys. Rev. Lett.} {\bf 104}, 250402 (2010)

\bibitem{dunne05} Dunne G. V., Heisenberg-Euler effective lagrangians: Basics and extensions, in Ian Kogan Memorial Collection
``From Fields to Strings: Circumnavigating Theoretical Physics", Vol 1, 445 (2005)

\bibitem{dunne09} Dunne G. V., New Strong-Field QED Effects at ELI: Nonperturbative vacuum pair production, \emph{Eur. Phys. J.} {\bf 55}, 327 (2009)

\bibitem{erde81} Erd\'{e}lyi A., Magnus W., Oberhettinger F. and Tricomi F. G.  \emph{Higher Transcendental Functions}, Vol. 1. (Krieger, New York, 1981)

\bibitem{elizalde94} Elizalde E., Odintsov S. D., Romeo A., Bytsenko A., and Zerbini S., \emph{Zeta  Regularization Techniques with Applications}, (World Scientific, Singapore, 1994)

\bibitem{elizalde09} Elizalde E., Odintsov S. D., and Saharian A. A., Repulsive Casimir effect from extra dimensions and Robin boundary conditions: From branes to pistons,
\emph{Phys. Rev.} {\bf D79}, 065023 (2009)



\bibitem{fucci10} Fucci G., and Avramidi I. G., On the gravitationally induced Schwinger mechanism,
In: ``Quantum Field Theory under the Influence of External Conditions'' (QFEXT09). Eds. Kimball A. Milton and Michael Bordag (Singapore: World Scientific, 2010), pp. 485-491

\bibitem{fucci11a} Fucci G., On the Hurwitz zeta function of imaginary second argument, \emph{J. Math. Phys.} {\bf 52}, 113501 (2011)

\bibitem{fucci11} Fucci G. and Kirsten K., Bose-Einstein condensation on product manifolds, \emph{J. Phys A: Math. and Theor.} {\bf 44}, 332002 (2011)

\bibitem{gavrilov95} Gavrilov S. P. and Gitman D. M., Vacuum instability in external fields, \emph{Phys. Rev.} {\bf D53}, 7162 (1995)

\bibitem{gelis15} Gelis F. and Tanji N., Schwinger mechanism revisited, \emph{Prog. Part. Nucl. Phys.} {\bf 87}, 1 (2016)

\bibitem{gilkey95} Gilkey P. B., \emph{Invariance Theory the Heat Equation and the Atiyah-Singer Index Theorem}, (CRC Press, Boca raton, 1995)

\bibitem{giudice} Giudice G., Rattazzi R., and Wells J. D., Quantum gravity and extra dimensions at high-energy colliders, \emph{Nucl. Phys. B}, {\bf 544}, 3 (1999)

\bibitem{haw77} Hawking S. W., Zeta function regularization of path integrals in curved space-time, \emph{Commun. Math. Phys.} {\bf 55}, 133 (1977)

\bibitem{heben09} Hebenstreit F., Alkofer R., Dunne G. V., and Gies H., Momentum signatures for Schwinger pair production in short laser pulses with sub-cycle structure,
\emph{Phys. Rev. Lett.} {\bf 102}, 150404 (2009)

\bibitem{heisen36} Heisenberg W. and Euler H., Consequences of Dirac’s theory of positrons, \emph{Z. Phys.} {\bf 98}, 714 (1936)

\bibitem{holst} Holstein B. R., Strong field pair production, \emph{Am. J. Phys.} {\bf 67}, 499 (1999)

\bibitem{kirsten01} Kirsten K., {\it Spectral Functions in Mathematics and Physics}, (CRC Press, Boca Raton, 2001)

\bibitem{kirsten09} Kirsten K. and Fulling S. A., Kaluza-Klein models as pistons, \emph{Phys. Rev.} {\bf D79}, 065019 (2009)

\bibitem{lewin} Lewin L., \emph{Polylogarithms and Associated Functions}, (North-Holland, New York, 1981)

\bibitem{niki} Nikishov A. I., Barrier scattering in field theory removal of Klein paradox, \emph{Nucl. Phys. B} {\bf 21}, 346 (1970)

\bibitem{orth11} Orthabera M., Hebenstreit F., and Alkofer R., Momentum spectra for dynamically assisted Schwinger pair production,
\emph{Phys. Lett. B} {\bf 698}, 80 (2011)

\bibitem{paris} Paris R. B. and Kaminski D., \emph{Asymptotics and Mellin-Barnes Integrals}, Encyclopedia of Mathematics and its Applications (No. 85),
(Cambridge University Press, Cambridge, 2001)

\bibitem{pop} Poppenhaegera K., Hossenfeldera S., Hofmannb S., and Bleichera M., The Casimir effect in the presence of compactified universal extra dimensions
\emph{Phys. Lett. B} {\bf 582}, 1 (2004)

\bibitem{rub} Rubakov V. A. and Shaposhnikov M. E., Extra space-time dimensions: towards a solution to the cosmological constant problem,
\emph{Phys. Lett. B} {\bf 125}, 139 (1983)

\bibitem{schu08} Schuetzhold R., Gies H., and Dunne G. V., Dynamically assisted Schwinger mechanism,  \emph{Phys. Rev. Lett.} {\bf 101}, 130404 (2008)

\bibitem{schwing51} Schwinger J., On gauge invariance and vacuum polarization, \emph{Phys. Rev.} {\bf 82}, 664 (1951)

\bibitem{vassi03} Vassilevich D. V., Heat kernel expansion: User's manual, \emph{Phys. Rep.} {\bf 388}, 279 (20030

\bibitem{whit90} Whittaker E. T. and Watson G. N. \emph{A Course in Modern Analysis}, (Cambridge University Press, Cambrige, 1990)

\end{thebibliography}
\end{document}